\documentclass{JINST}
\usepackage{multirow} 
\usepackage[figuresright]{rotating}
\pdfoutput 1
\title{A prototype liquid Argon Time Projection Chamber for the study of UV laser multi-photonic ionization} 

\author{ B.~Rossi\thanks{Corresponding author.}, I.~Badhress, A.~Ereditato,  S.~Haug,  R.~H\"anni, M.~Hess, S.~Jano\^{s}, F.~Juget,  I.~Kreslo,  S.~Lehmann, P.~Lutz, R.~Mathieu, M.~Messina, U.~Moser, F.~Nydegger,  H.U.~Sch\"utz, M.S.~Weber, M.~Zeller \\
  \llap{}Laboratory for High Energy Physics (LHEP)\\
  Centre for Research and Education in Fundamental Physics,\\
  University of Bern,\\
  Switzerland \\
  E-mail: \email{biagio.rossi@lhep.unibe.ch}
}

\abstract{This paper describes the design, realization and operation of a prototype liquid Argon Time Projection Chamber (LAr TPC) detector dedicated to the development of a novel online monitoring and calibration system exploiting UV laser beams. 
In particular, the system is intended to measure the lifetime of the primary ionization in LAr, in turn related to the LAr purity level. 
This technique could be exploited by present and next generation large mass LAr TPCs for which monitoring of the performance and calibration plays an important role.
Results from the first measurements are presented together with some considerations and outlook.}


\begin{document}
\section{Introduction}
Liquid Argon Time Projection Chamber (LAr TPC) devices were originally proposed and developed for the detection of ionizing events \cite{[1]}. The technology allows for uniform and high resolution imaging of massive detector volumes. The operating principle of the LAr TPC is based on the undistorted tracks of ionizing electrons in highly purified LAr by a uniform electric field over distances of the order of metres \cite{[2]}. Imaging is provided by wire planes or other read-out devices placed at the end of the drift path. 
The drifting electrons are collected by the outermost wire plane which gives position and calorimetric information. Additional planes with different orientation are positioned in front of the collection plane and record the signal induced by the passage of the drifting electrons.   
This provides different projective views of the same event and therefore allows space point reconstruction. The third coordinate is given by the measurements of the drift time given by the time interval between the passage of the ionizing particle in the active volume ($t_0$) and the arrival of the drifting electrons on the wire planes. The $t_0$ can come from the detection of the scintillation light of the liquid Argon by means of photomultipliers, or from an external source.

The particle momentum of an incoming particle is inferred by the measurement of its multiple scattering \cite{[nrg_scattering]}, while the detection of the local energy deposition provides particle identification. The total energy reconstruction of the track is performed by charge integration within the detector volume, being the device a full-sampling, homogenous calorimeter.

The purity of the liquid Argon is a key ingredient to achieve imaging over long drift distances. A purity corresponding to less than 0.1~ppb of electro-negative elements such as Oxygen has to be achieved in order to allow for electron lifetimes of milliseconds and, hence, drift distances of meters, as envisioned for large scale applications of this technique  \cite{[7],[8],[9],[10],[13]}. This can be obtained by means of commercial purification cartridges and by the re-circulation of the Argon.

In this paper we describe the design, realization, operation and test of a LAr TPC prototype used as a test facility for novel read-out devices, front-end electronics, DAQ and HV systems, as well as of cryogenic purification systems, and for the study of event reconstruction and pattern recognition software. 
In addition, we report on our plan for the study of multi-photon liquid Argon ionization by means of UV laser light. The goal is to develop a practical and efficient new method for the general monitoring of large mass LAr TPC detectors and of the purity of the LAr for use in the future applications of the technique. 
The laser beam can also be used to calibrate the TPC. This is based on the possibility of producing tracks inside the LAr detector volume that are similar to those created by ionizing particles and of which one knows (a priori) the ionization yield, the position, the arrival time, and that the energy released is not affected by Landau fluctuations.

The prototype described in this paper is part of a R\&D project we are conducing in Bern envisioning the realization of LAr TPC detectors of increasing size, combining our efforts with additional and complementary studies conducted at ETHZ \cite{[LEM]}. First results have been published on the study of a TPC we have built and operated employing a mixture of LAr and Nitrogen \cite{[LN]}. 

The above mentioned need for achieving very long electron drift paths is one of the key issues of our program. The feasibility of long drift distances is being studied experimentally by the construction of a 5 metres long-drift LAr TPC (ARGONTUBE) in our laboratory. 
%
\section{UV laser ionization in noble liquids}
\indent

Laser ionization occurs as a multi-photon process through the simultaneous absorption of two or more photons via virtual states in a medium. This process requires high photons flux from pulsed lasers. 
Bound electrons may absorb several laser photons simultaneously to excite the atom to high energy levels or even to overcome the ionization potential. 
The ionization potential for Argon (in gas phase) is I$_\mathrm{p}$ = 15.76~eV \cite{[Schmidt]}. The photon energy at the wavelength used for the measurement reported here ($\lambda$ = 266~nm) is E$_{\mathrm{ph}}$ = 4.66~eV.
Thus, a minimum number of four photons is required to produce ionization.  The absorption of multiple photons makes this process highly non-linear ($4^{\mathrm{th}}$ order for Ar); its cross-section strongly diminishes with the order of the process, thus making its effectiveness largely depending on the laser wavelength. 

Interaction of laser light with matter in the liquid phase has been comparatively less studied with respect to gas and solid phase. Nevertheless, some important features are well understood and reported in \cite{[Schmidt]}. In particular, it is expected that the ionization potential is approximately a few tenths of eV below 14~eV. Thus, we expect that three photons are required to ionize liquid Argon.

UV laser ionization of LAr has been shown by a previous experiment \cite{[Sun]}. However, the order of the process has not yet been determined. With our setup we intend to study the electron yield as a function of the laser intensity allowing to infer the order of the process, hence gaining information about the laser-liquid interaction process.

\section {Description of the detector and of the related systems}
\indent

\begin{figure}
\center\includegraphics[width=14cm]{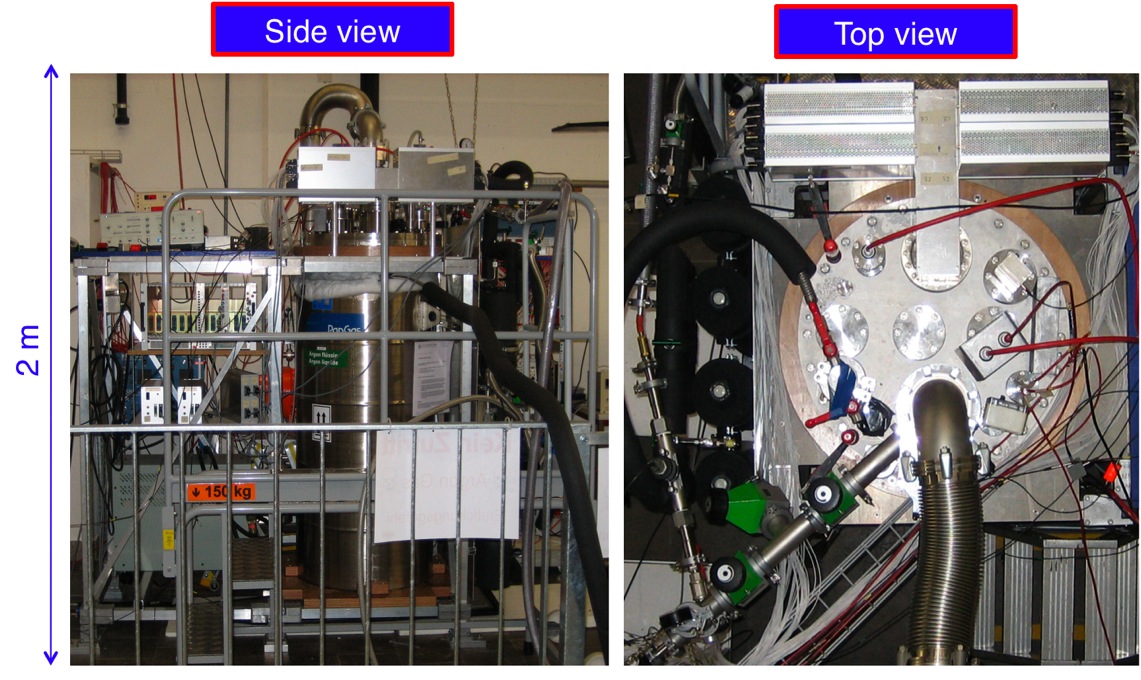}
\caption{Photograph of the experimental setup.} 
\label{fig:dewar}
\end{figure}

\begin{figure}
\center\includegraphics[width=5cm]{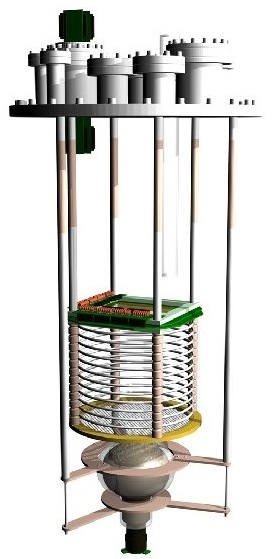}
\caption{Artistic view of the detector. The top flange, the TPC, the PMT and the realtives supports are shown.} 
\label{fig:dewar2}
\end{figure}

\begin{figure}
\center\includegraphics[width=5cm]{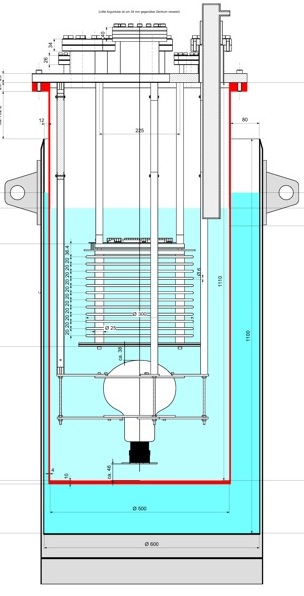}
\caption{Schematic drawing of the detector set-up. The TPC, mounted with the drift coordinate vertically.
The PMT and the relative supports are housed in a cylindrical tube filled with pure Argon. The external vessel contains un-purified Argon. } 
\label{fig:dewar5}
\end{figure}

The detector (figure~\ref{fig:dewar} and~\ref{fig:dewar2}) consists of a TPC housed in a tube filled with purified liquid Argon, in thermal contact with a bath of liquid Argon (figure~\ref{fig:dewar5}). The TPC is complemented by an electronic read-out and data acquisition system (DAQ), a liquid Argon re-circulation and purification system, a photomultiplier (PMT), a series of ancillary equipments for monitoring and control, and a UV-laser with the related optics to allow the study of multi-photon liquid Argon ionization. The various components are described in detail in the following sections.

%
\subsection{Cryogenic vessels and vacuum flanges}
\begin{figure}
  \center\includegraphics[width=14cm]{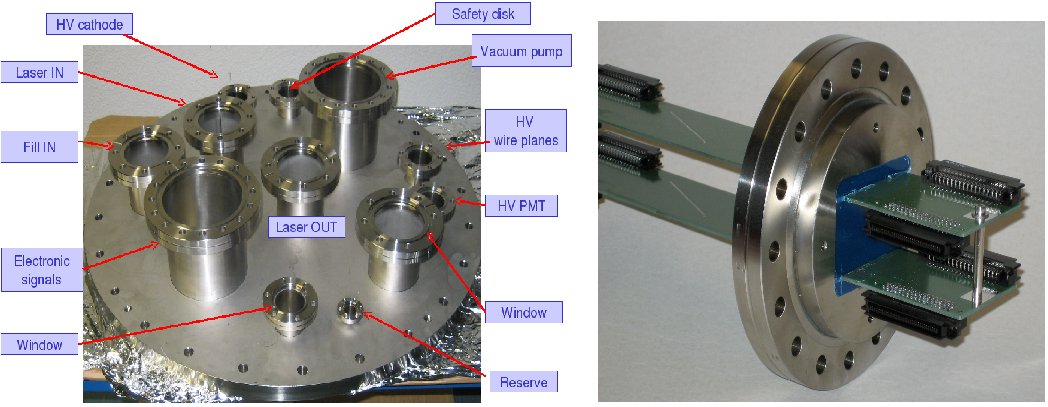}
  \caption{Left: photograph of the vessel closing flange. Right: signal feed-through flange.} 
  \label{fig:flange}
\end{figure}

\indent
The TPC detector is contained in a stainless-steel cylindrical tube. The latter measures 50~cm in diametre and 110~cm in height, for an inner volume of about 200~litres. The tube is placed in thermal contact with a bath of liquid Argon and is closed by a stainless-steel flange. Vacuum tightness is guaranteed by an Indium O-ring positioned in a circular 2~mm groove on the top part of the vessel. A series of additional smaller flanges are positioned on the top flange (figure~\ref{fig:flange}). They include feed-throughs needed to evacuate the inner volume of the vessel, to fill it with liquid Argon, to read-out the detector signals, to supply the high-voltage, to house the pressure monitors, level metre and safety valves, and to provide the transport of the UV laser beams to the inner part of the TPC. 

\begin{figure}
  \center\includegraphics[width=7cm]{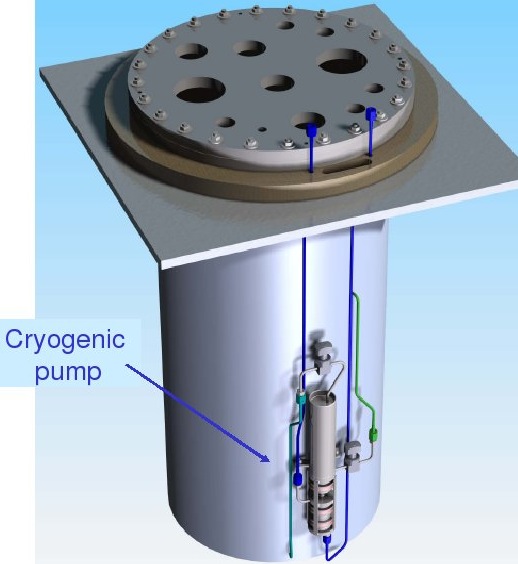}
  \caption{Artistic view of the cryogenic re-circulation pump and the filter cartridge.} 
  \label{fig:rec_pump}
\end{figure}

\begin{figure}
  \center\includegraphics[width=3cm]{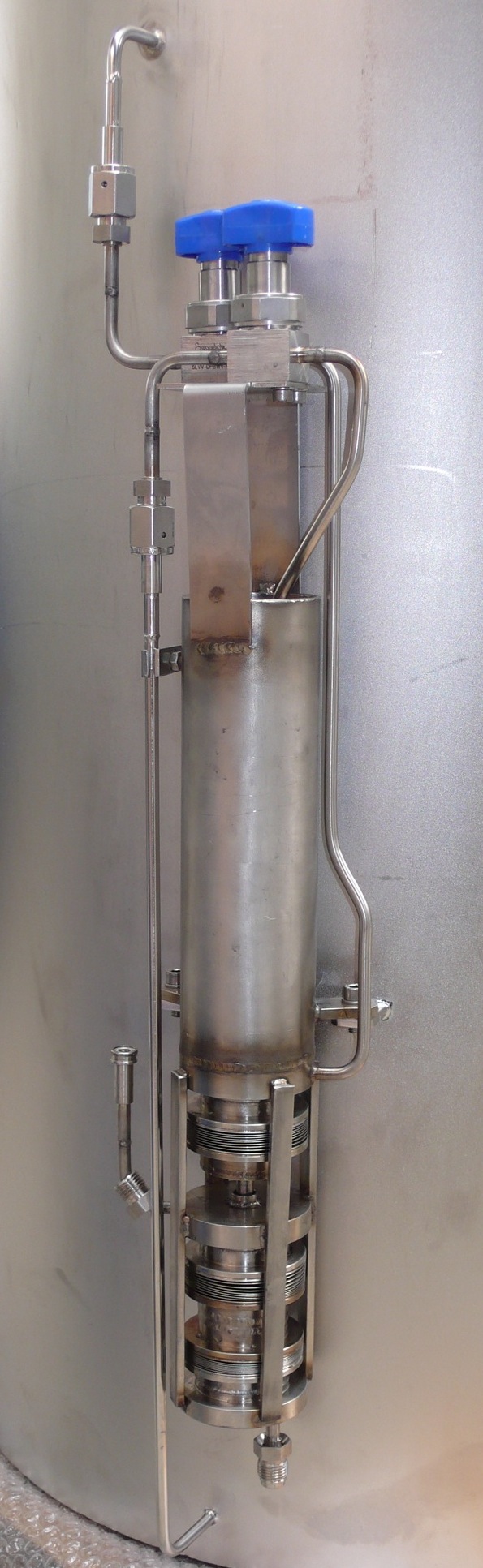}
  \caption{Photograph of the cryogenic re-circulation pump.} 
  \label{fig:rec_pump_photo}
\end{figure}

The major sources producing impurities in the liquid Argon are the out-gassing detector components which are not in direct contact with the liquid (flanges, feed-throughs, cables) and have, therefore, a higher temperature. A cryogenic mechanical pump (figure~\ref{fig:rec_pump} and~\ref{fig:rec_pump_photo}) immersed in the bath of liquid Argon allows to keep the level of impurities at a low level by forcing the liquid Argon re-circulation through an Oxygen filter and a water filter mounted in series (Oxysorb, Hydrosorb~\cite{[15]}). The re-circulation speed is $\sim$1~l/hour in the present configuration. A new system able to provide up to 10-30~l/hour recirculation speed will be operational for future measurements. 
%
\subsection{TPC, HV system and read-out electrodes}
\begin{figure}
  \center\includegraphics[width=14cm]{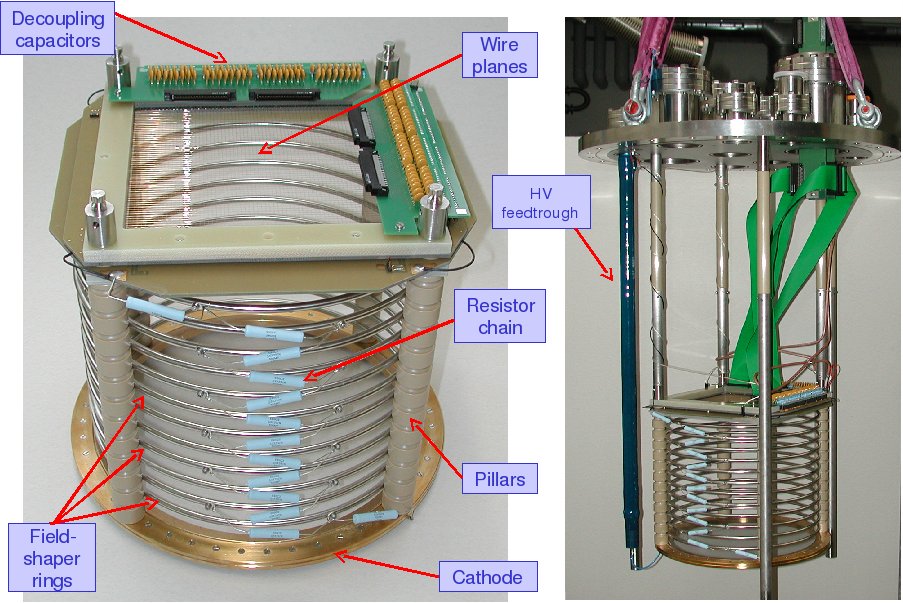}
  \caption{Left: photograph of the TPC; some of the main components are indicated. Right: the TPC after assembly ready for data taking. The HV feedthrough is also visible.} 
  \label{fig:TPC}
\end{figure}

\indent
The TPC has the shape of a parallelepiped inscribed in a cylinder delimited by the field shaping rings, sustained by four cylindrical bars made of stainless-steel and PEEK, fixed to the closing flange (figure~\ref{fig:TPC}). PEEK is an organic polymer that has been chosen for its good thermo-mechanical properties at the temperature of liquid Argon. 
The TPC cathode is a metallic mesh held by two cylindrical frames (320~mm diametre). The anode (read-out electrodes) consists of two parallel wire planes spaced by 3~mm. Each plane is made of 64 Be-Cu alloy wires, 0.125~mm in diametre, with 3~mm pitch and 200~mm length. The first wire plane works in induction mode while the second plane collects the drifting electrons (collection mode). The wires of the induction plane run orthogonal to the ones of the second plane. 

\begin{figure}
\center\includegraphics[width=15cm]{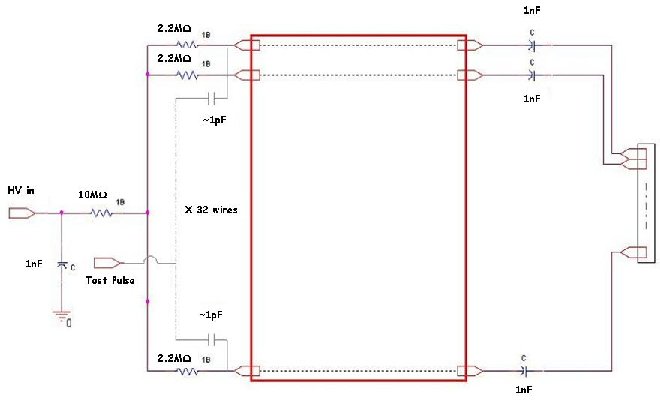}
\caption{Electric scheme of the wire planes.} 
\label{fig:wire_electricscheme}
\end{figure}

The wires are soldered on a Vetronite frame that also holds the test-pulser printed lines and the high-voltage distribution board. The latter hosts the decoupling capacitors connecting the sensing wires to the front-end electronics and the connector for the signal cables. The electric scheme of the wire planes is shown in figure~\ref{fig:wire_electricscheme}. 
The distance between cathode and anode can easily be modified for different experimental conditions in the range 5~mm to 800~mm. In the tests reported here we have chosen it equal to 260~mm.
In order to make the drifting electric field uniform, 12 field-shaping electrodes made of stainless-steel rings of 300 mm diametre and 6~mm cross-section are placed between anode and cathode at equal spacing (figure~\ref{fig:TPC}). Each ring is connected to the previous one via 100 M$\Omega$ resistors, working as high-voltage dividers. The first and the last ring are also connected to the cathode and to the induction wire plane via a 100~M$\Omega$ resistor, respectively. 

The high-voltage for the drift field is brought to the cathode by a feed-through able to withstand -30~kV. 
Feeding such a high voltage from room temperature into the liquid is a challenge. The existing technology relies on the accurate matching of the thermal expansion factors of the insulator and of the metal coaxial electrodes. We followed an approach that is much less sensitive to thermal expansion. The HV feedthrough has a coaxial structure with the central electrode consisting of multy-thread twisted wire and the outer (ground) electrode made of fine wire mesh, similar to the one of flexible coaxial cables. The insulator used is a polyurethane resin\footnote{Arathane CW5620.} developed to withstand temperatures as low as 77~K  without loosing its insulating properties (25~kV/cm). The mesh and the twisted central conductor have high elasticity, so that the contraction at low temperature does not impose mechanical stress on the insulator. The feedthrough has demonstrated high stability after multiple heating-cooling cycles.

A typical field of 1~kV/cm is used to run the TPC. 
The transparency of the induction plane is 100\% if
\begin{equation}
\frac{E_{\mathrm{col/ind}}}{E_{\mathrm{cat/ind}}}>\frac{1+\rho}{1-\rho}
\label{eq:transparency}
\end{equation}
where $E_{\mathrm{col/ind}}$ is the electric field between collection and induction plane, $E_{\mathrm{cat/ind}}$ is the field between the cathode and the induction plane, and 
\begin{equation}
\rho=\frac{2\pi r}{p}
\label{eq:x}
\end{equation}
with $r$ the wire radius and $p$ the pitch. With our specific choice for $r$ and $p$ we obtained $E_{\mathrm{col/ind}}>1.4E_{\mathrm{cat/ind}}$. Consequently, in order to ensure transparency, a field of 2.0~kV/cm is applied between the two wire planes; this is obtained by biasing the first induction plane to ground and the collection plane to +600~V.

The field shielding inefficiency $\sigma$ of the induction plane is defined by
\begin{equation}
\sigma=\frac{p}{2\pi d_{\mathrm{ind/cat}}}\mathrm{ln}\Big{(}\frac{p}{2\pi r}\Big{)}
\label{eq:inefficiency}
\end{equation}
where $d_{\mathrm{ind/cat}}$ is the distance between induction plane and cathode. The resulting inefficiency is about 3\%. It has to be noted that Eq. \ref{eq:transparency} and \ref{eq:inefficiency} are valid only for a grid of equally spaced parallel wires \cite{[Shield1], [Shield2]}.

\begin{figure}
  \center\includegraphics[width=9cm]{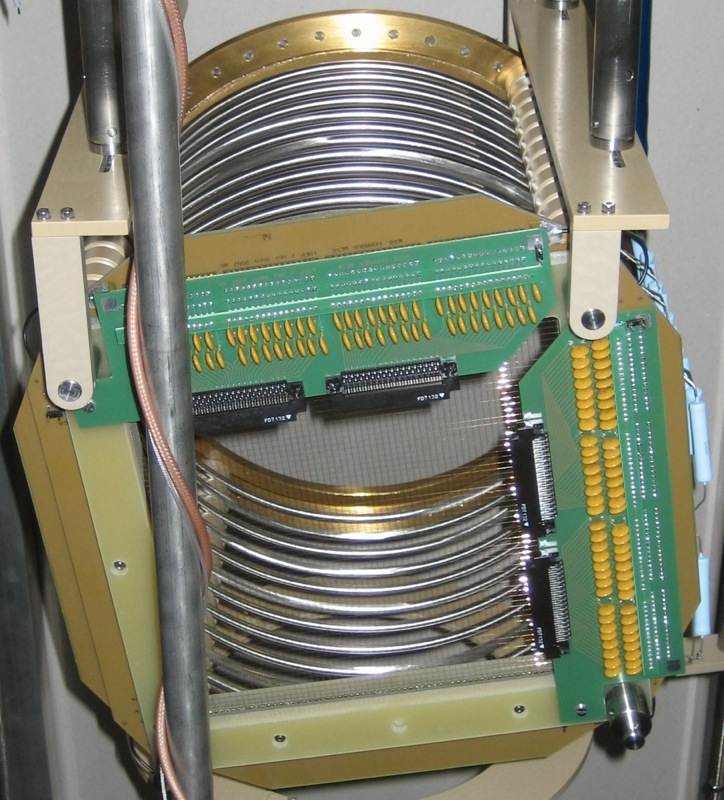}
  \caption{Photograph of the TPC in the configuration used to run the laser beam. The drift direction is horizontal, with induction (collection) wires running at an angle of 7 degrees relative to the vertical (horizontal) direction.}
  \label{fig:TPC_h}
\end{figure}

The TPC can be positioned inside the vessel in two different orientations: with the electrodes parallel to the ground for a vertical drift (figure~\ref{fig:TPC}), and with the electrodes orthogonal to the ground for a horizontal drift (figure~\ref{fig:TPC_h}). The latter configuration has been used for the UV laser ionization measurements.  

\subsection{UV laser and optics}
\label{par:laser}
\begin{figure}
\begin{center}
\includegraphics[width=13cm]{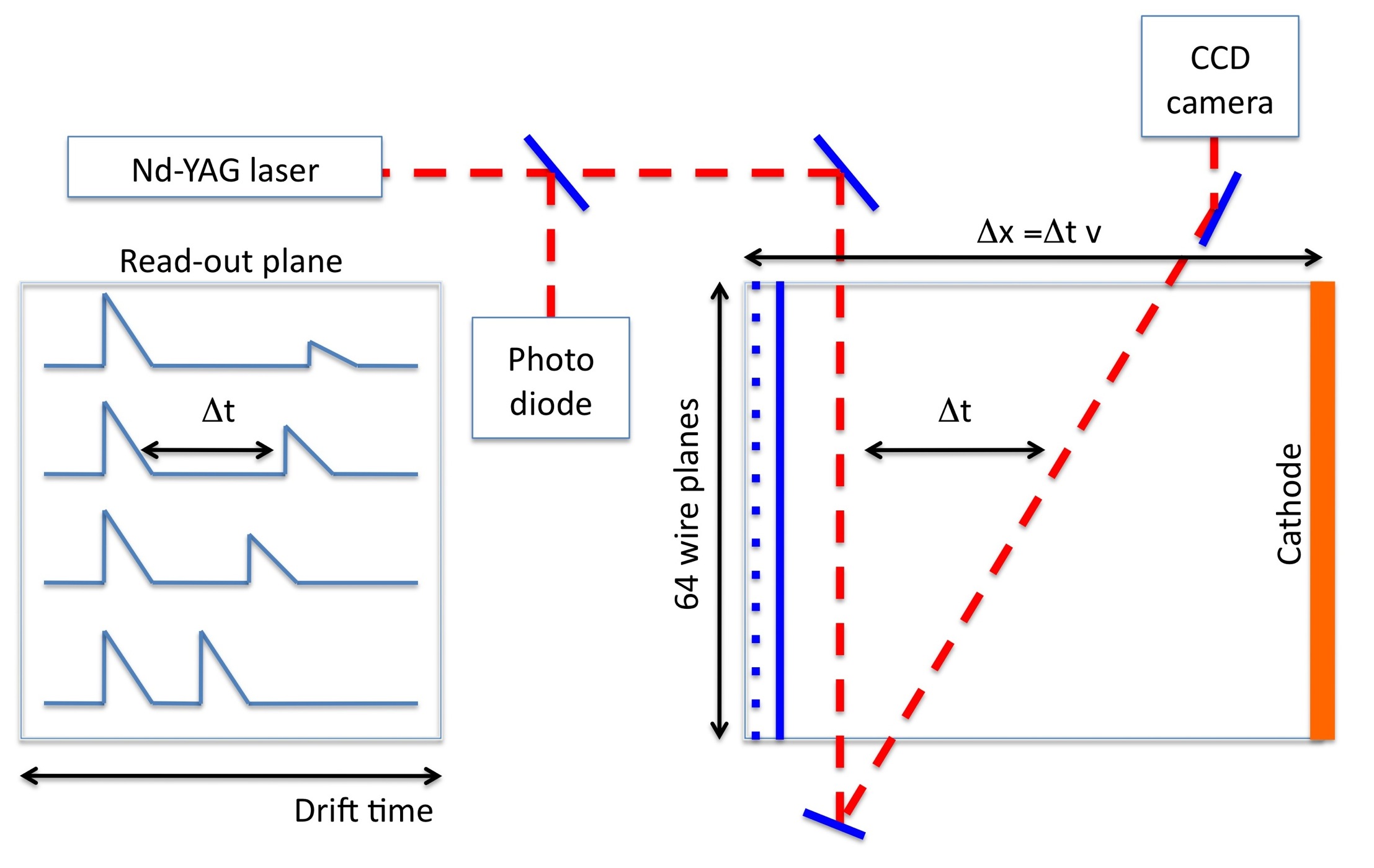}
\end{center}
\caption{\small Schematic view of the laser beam path and the read-out response.} 
\label{fig:laser_path}
\end{figure}

\begin{figure}
\begin{center}
\includegraphics[width=13cm]{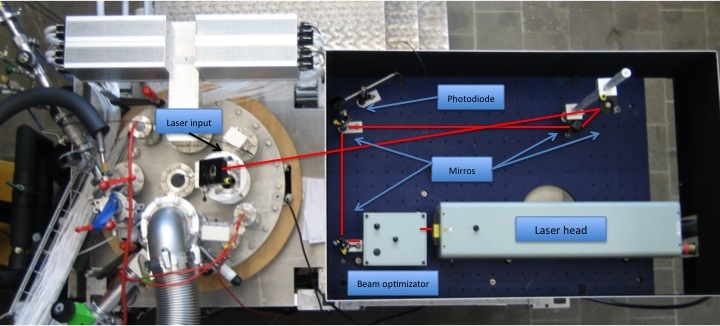}
\end{center}
\caption{\small Photograph of the experimental setup including the laser beam. Red lines show the beam path before being steered into the detector.} 
\label{fig:laser_setup}
\end{figure}

\begin{figure}
\begin{center}
\includegraphics[width=9cm]{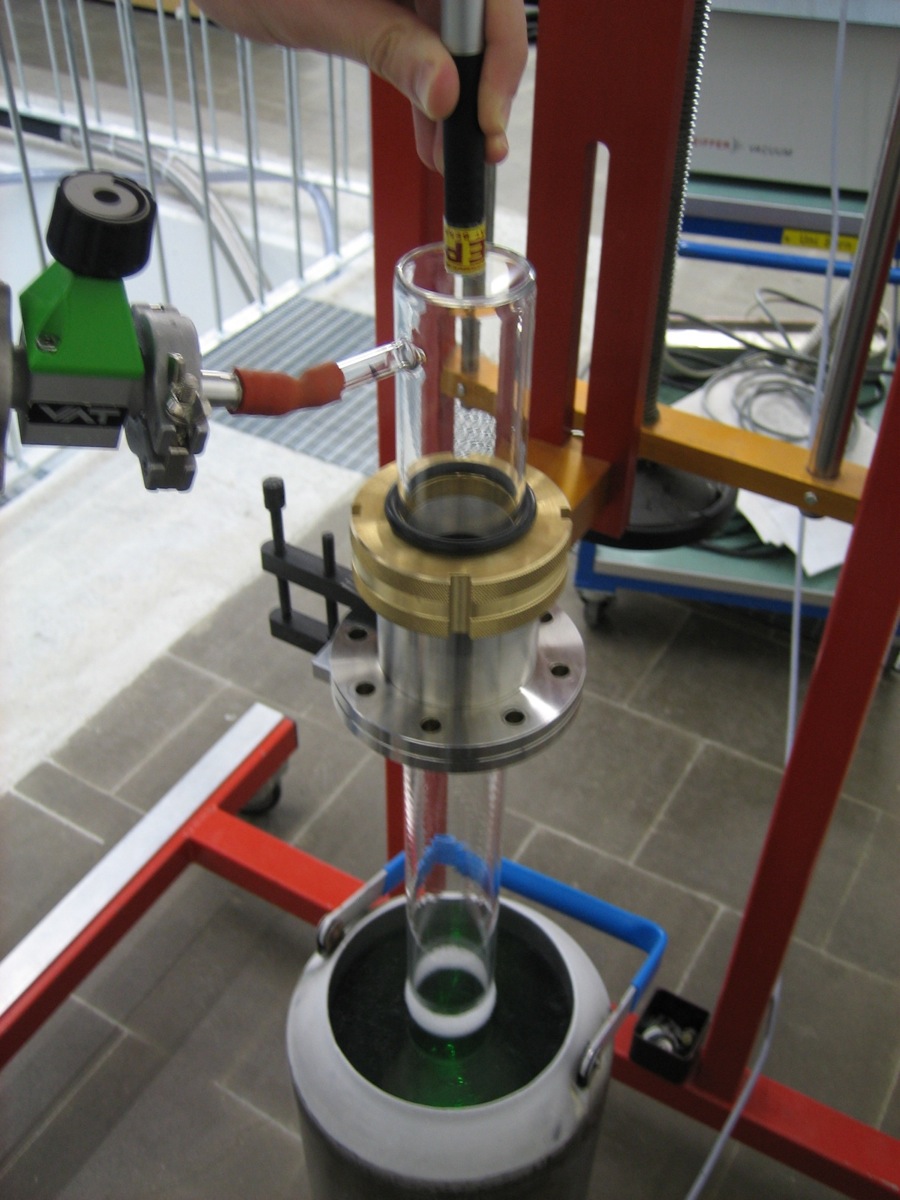}
\end{center}
\caption{\small Photograph showing the glass tube used as an optical feed-through during the termo-mechanical test.} 
\label{fig:optical_feed}
\end{figure}

\indent
For our measurements we used a pulsed ultraviolet Nd-YAG laser with harmonic generators\footnote{Continuum - model Surelite I-10.}. Its main characteristics are reported in Table~\ref{tab:laser}. In order to ionize liquid Argon we use the $4^{\mathrm{th}}$ harmonic with a wavelength of 266~nm corresponding to 4.66~eV photons. A minimum of three photons (14.01~eV) should be sufficient to ionize LAr.
\begin{table}[ht]
\caption{$4^{\mathrm{th}}$ harmonic UV laser specifications.}
\centering
\begin{tabular}{  c c c }
\hline\hline
wavelenght (nm) & max repetition rate (Hz) &  max energy (mJ)   \\ 
         266                &            10                             &         82                    \\ 
pulse width (ns)   & rod diameter (mm)           & divergence (mrad) \\
      4-6                    &               6                              &  0.6                          \\

\hline
\end{tabular}	
\label{tab:laser}
\end{table}
Figure~\ref{fig:laser_path} shows the planned arrangement and the path of the laser beam used for the tests. Both energy and pulse length measurements of the laser beam are done before the beam is actually steered into the chamber. The light is fed into the TPC inner volume by passing through a system of mirrors (figure~\ref{fig:laser_setup}) and an optical feed-through. The latter consists of a 600~mm long glass tube of 60~mm diametre with two quartz windows held by a stainless steel flange. A brass ring presses three O-rings to preserve the tightness of the main dewar. The design of the optical feed-through is a novel concept that we developed. The tube is partially immersed in liquid Argon (figure~\ref{fig:optical_feed}) to prevent beam deformations during the passage through the gas-liquid interface. 

For the measurements reported here the laser beam is dumped after passing through the active volume of the TPC. For the future, we plan to have the beam passing two times in the active volume: first vertically, as done now, and then at an angle with respect to the drift coordinate. After the second passage, the beam can be steered out of the dewar by means of a mirror and another optical feedthrough. The extraction of the laser beam from the TPC allows for the monitoring of its spot size with a CCD camera, needed for the measurement of the cross section of the ionization process.

The measurement of the pulse width is performed by means of a fast photodiode\footnote{Thorlabs DET 10 A.} read-out by a 2~GS/s and 1~GHz bandwidth oscilloscope. The measurement of the energy is presently performed with a photodiode and an energy meter\footnote{GEANTEC - SOLO 2.}, while for future tests this will be done with both the photodiode and CCD camera.  

The laser beam is used to measure the electron drift velocity and diffusion, the electron-ion recombination, the detector intrinsic energy resolution (no Landau fluctuation), the space resolution (no multiple scattering), as well as for the precise monitoring of the LAr purity. Results are shown in Section \ref{sec:data_taking} and \ref{par:results}.
%
\subsection{Read-out, data acquisition and trigger}
\begin{figure}
\center\includegraphics[width=12cm]{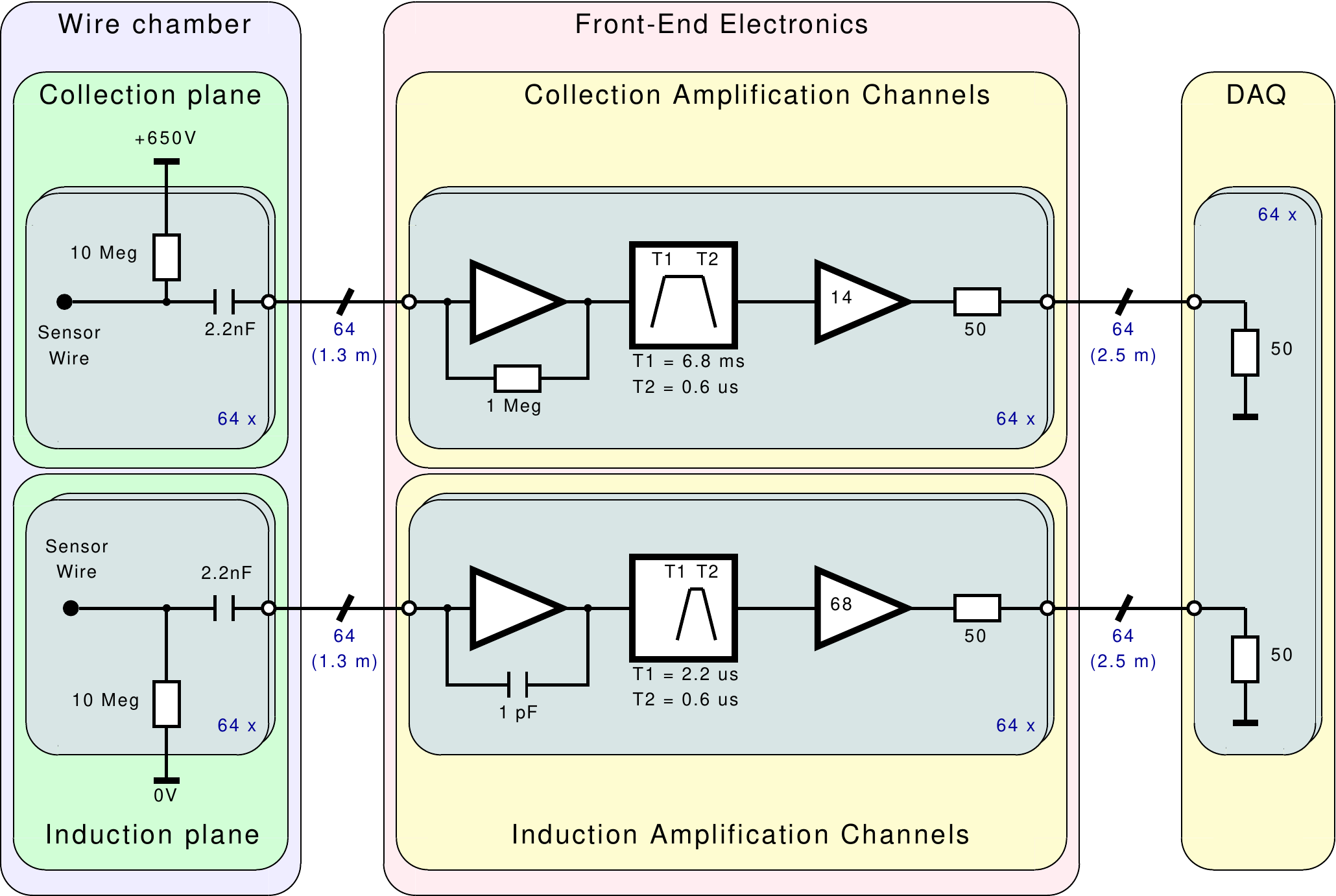}
\caption{Diagram of the Front-End electronics box.} 
\label{fig:E_chain}
\end{figure}

\indent
The electronic read-out chain (figure~\ref{fig:E_chain}) is composed of three basic units. The first one is a decoupling board plugged onto the wire planes, receiving the signals from the TPC wires and passing them to the Front-end boxes. The second is a front-end box, that houses the signal pre-amplifiers, a linear regulator and test pulse inputs. The third unit is a digital board with the circular digital buffers and the VME\footnote{ANSI/IEEE 1014-1987.} interface. 
 
\begin{figure}
\center\includegraphics[width=12cm]{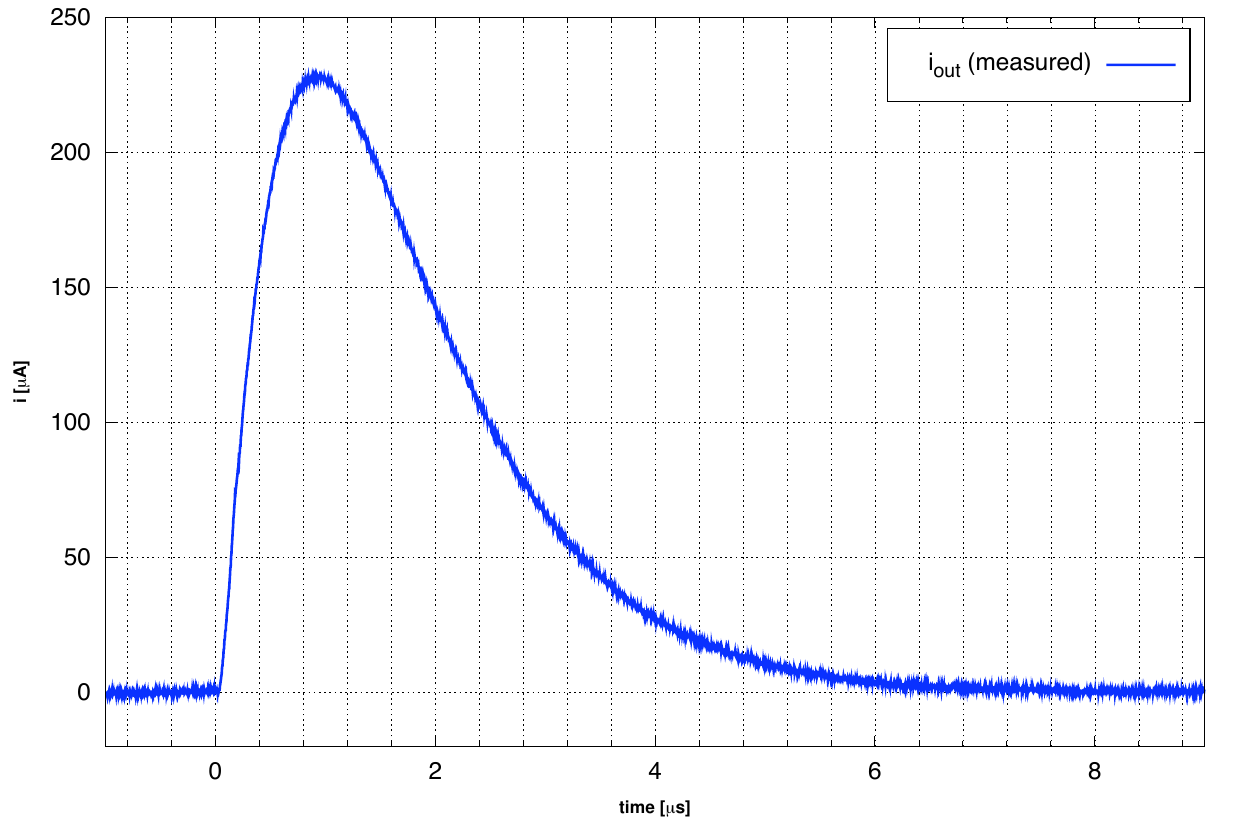}
\caption{Response of the Collection plane preamplifiers to a 30~nA, 130~ns test pulse. This pulse simulates the signal induced by a track with a large angle ($\sim$ 90 degrees) with respect to the drift direction.} 
\label{fig:short_pulse}
\end{figure}

\begin{figure}
\center\includegraphics[width=12cm]{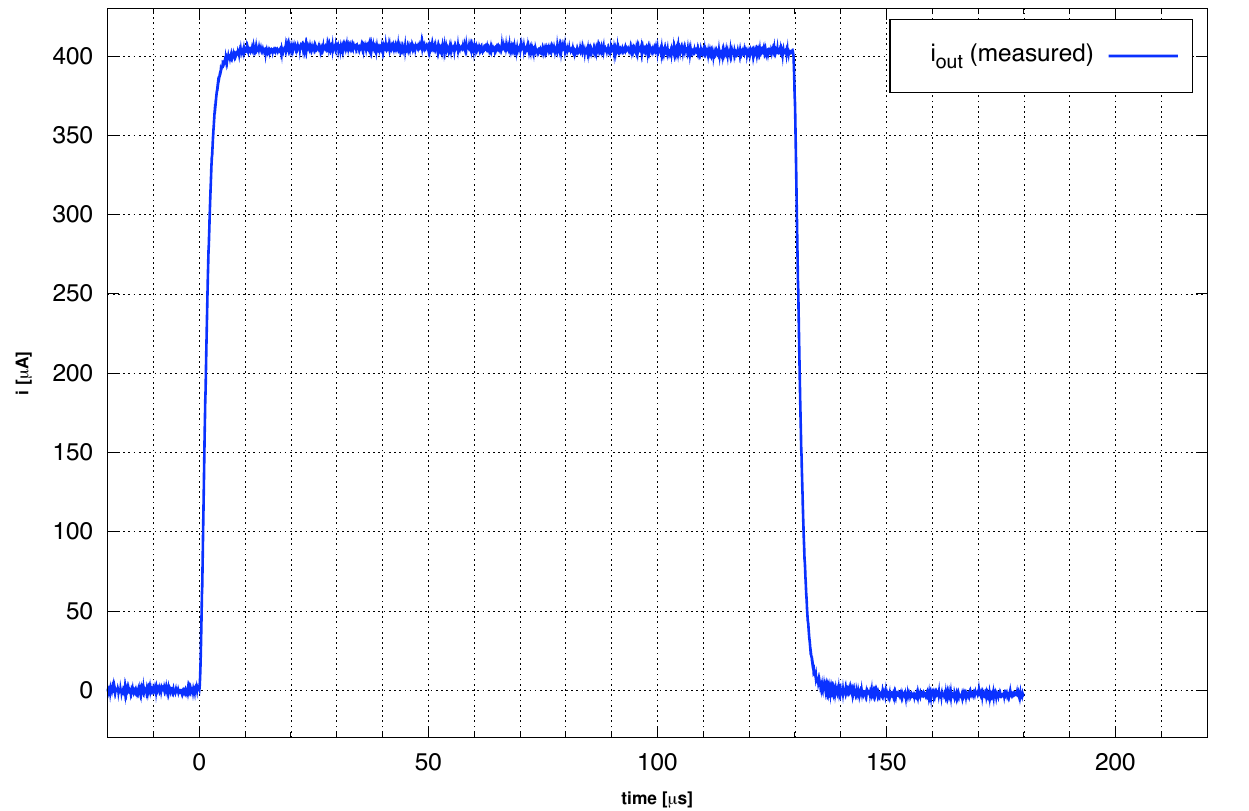}
\caption{The same as the previous figure, but for a 3~nA, 130~$\mu$s test pulse. This pulse simulates the signal induced by a track with a small angle with respect to the drift direction.} 
\label{fig:long_pulse}
\end{figure}

The decoupling boards of each wire plane are connected via flat ribbon cables  and feed-through flanges to the four read-out front-end boxes. Each front-end box houses 16 pre-amplifier cards with two channels each. The Induction (Collection) plane is read-out by preamplifiers working in charge integration (current) mode. The charge induced on the Induction plane by the drifting electrons is integrated and shaped to maximize the signal to noise ratio. 
The requirements to the Collection plane for the simultaneous track and energy reconstruction do not allow the use of shapers in order to reduce the noise. 
A typical signal is a few $\mu$s long, but it may become much longer ($\sim$130~$\mu$s) if the track has a small angle with respect to the drift direction. This implies that the low frequency components must be amplified without distortion. Therefore, the Collection plane preamplifiers are set to work in current mode with a very large bandwidth. In figure~\ref{fig:short_pulse} the response of the Collection plane preamplifiers to a 130~ns current pulse is shown. This simulates the signal from a track with a large angle ($\sim$90~degrees) with respect to the drift direction. Conversely, in figure~\ref{fig:long_pulse} we present the response of the preamplifiers to a 130~$\mu$s current pulse, simulating the signal from a track with a small angle with respect to the drift direction. The output waveform follows without distortion the input current.

The 16 digital boards\footnote{CAEN V1724.} are housed in a VME crate\footnote{CAEN VME8101 LC 6U.} and read-out by a VME to PCI optical bridge\footnote{CAEN V2718 Controller.}, with a maximal data flow of 70~MB/s. The ADCs have 8 channels, 14~bit resolution and 100~MS/s maximum sampling rate. A SRAM memory (512~ksamples/ch) with independent read-write access divided in buffers of programmable size (512$\div$512~ksamples) is installed on the boards. For our purposes the buffer size was chosen equal to 1024~samples/ch corresponding to 262~kB per event. With this configuration a hardware pipeline of 512 events is available. When a trigger occurs, the FPGA writes the first event and freezes the current buffer. The ADCs continue conversion and a new event is written into the next buffer segment without dead time. 
The handshaking between circular buffer and readout software is done using the VME bus error line. The latter is active unless converted data are available.
The DAQ software includes the following components: the {\it Graphic User Interface} (GUI) used to set-up run conditions, ADC settings, and to run the acquisition and the monitoring of the events, the {\it VME connection} to handle the VME Controller and transfer data from the ADCs to the disk storage, and the {\it on-line monitoring} of the events.

\begin{figure}
\center\includegraphics[width=12cm]{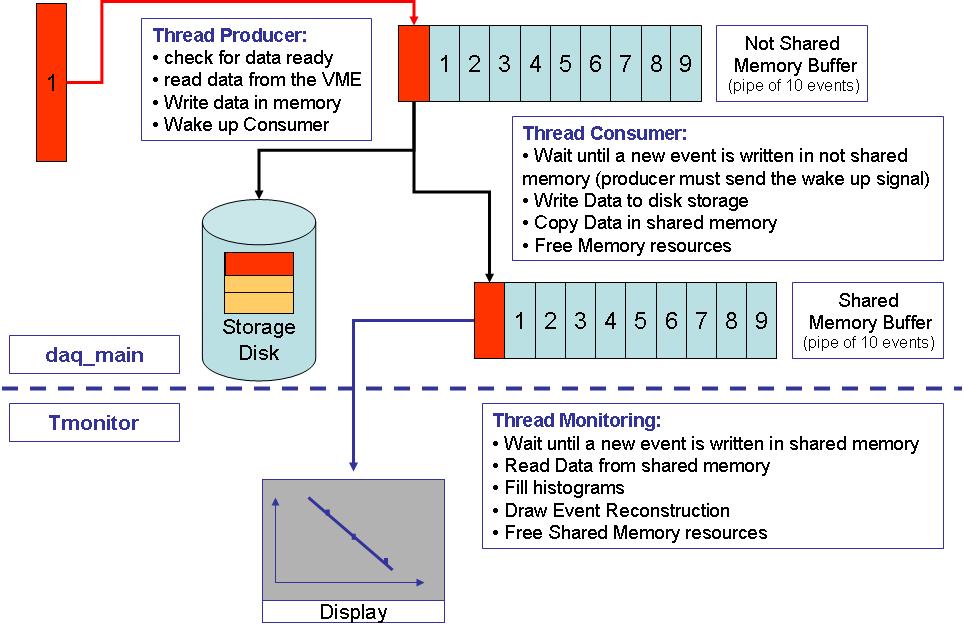}
\caption{Schematic view of the data acquisition software.} 
\label{fig:DAQ_schema}
\end{figure}

A schematic view of the data acquisition software is presented in figure~\ref{fig:DAQ_schema}. A software (SW) {\it not shared memory buffer} able to contain up to 10 events used as a data pipeline, and the related semaphores that regulate the access to the memory itself by different processes, are defined in the {\it VME connection} executable. 
A {\it memory buffer shared} between {\it VME connection} and {\it on-line monitoring} executables, containing a copy of the raw data, is also created. Its aim is giving access to raw data for monitoring purposes without introducing dead time in the DAQ data transfer. 

Two main threads (producer and consumer) are defined and used to implement parallelism. 
The {\it producer} continuously checks the bus error line for available data. When they are available the reading of the first event from the ADCs' SRAM and the transfer to the SW memory buffer is performed. Then, the {\it producer} wakes up the {\it consumer}, acting on a semaphore, and checks again for new data ready.
The {\it consumer} sleeps until a new event is written in the SW memory buffer, reads it and writes it to the disk storage. The same event is copied in to the {\it shared memory buffer}, making it accessible to the {\it on-line monitoring} executable. In the end, the memory resources in the {\it un-shared buffer} are freed. 
A thread of the {\it on-line monitoring} executable reads the {\it shared buffer} when a new event is present. Filling and drawing of histograms is then performed and the resources of the shared pipeline are freed. 

The usage of different executables makes the software more stable. For example, the {\it on-line monitoring} can stop without affecting the data acquisition. Unless the writing speed on the disk storage is slower than the data flow from the VME controller to the PC, no dead time is introduced. 

The executables are written in C/C++ by using the open-source Qt graphic libraries\footnote{www.trolltech.com} and the ROOT\footnote{root.cern.ch} framework graphic and analysis packages. 
It is architecture independent (tested on 32-bit and 64-bit linux machines) and in our setup it runs on a four multi-core processor machine.  

The TPC is equipped with a photomultiplier (PMT) immersed in LAr (figure~\ref{fig:dewar2}), which generates a fast signal (4~ns) from the scintillation light in the liquid Argon, used as trigger for the DAQ system. The trigger signal from the PMT gives the time information ($t_0$) necessary for the reconstruction of the third spatial coordinate of the TPC. The PMT is a $8"$  Hamamatsu R-5912-02MOD that provides good performance at LAr temperature.
The spectral emission of the scintillation light of LAr has its maximum value at 128~nm. Light with such a wavelength is absorbed by the glass window of the PMT and, furthermore, does not match the absorption spectrum of the PMT photocathode. To overcome this problem the PMT glass window was coated with a wavelength shifter made of Tetraphenylbutadiene (TPB). The TPB coating was applied by means of a bath in a solution of TPB, polystyrene and chloroform \cite{[16]}. 

The output signal of the PMT is sent to a shaping amplifier with $1~\mu$s shaping time to be acquired by the ADC. It is also sent to a discriminator to generate a logical signal used as trigger for the DAQ.  The threshold of the discriminator was chosen to select clean cosmic-ray events with high-energy muons traversing the full sensitive volume of the detector. 

The DAQ can also run with a $t_0$ signal given by the pulsed laser. The buffer size has been chosen to provide 1024 samples (each 210~ns long), enough to record a full drift length with the TPC electric field at 1~kV/cm. Events are written in raw-data format without any zero suppression. Both noise conditions and signal calibrations are periodically checked with test-pulses.

\section{Detector operation}
\indent
\begin{figure}
\center\includegraphics[width=13cm]{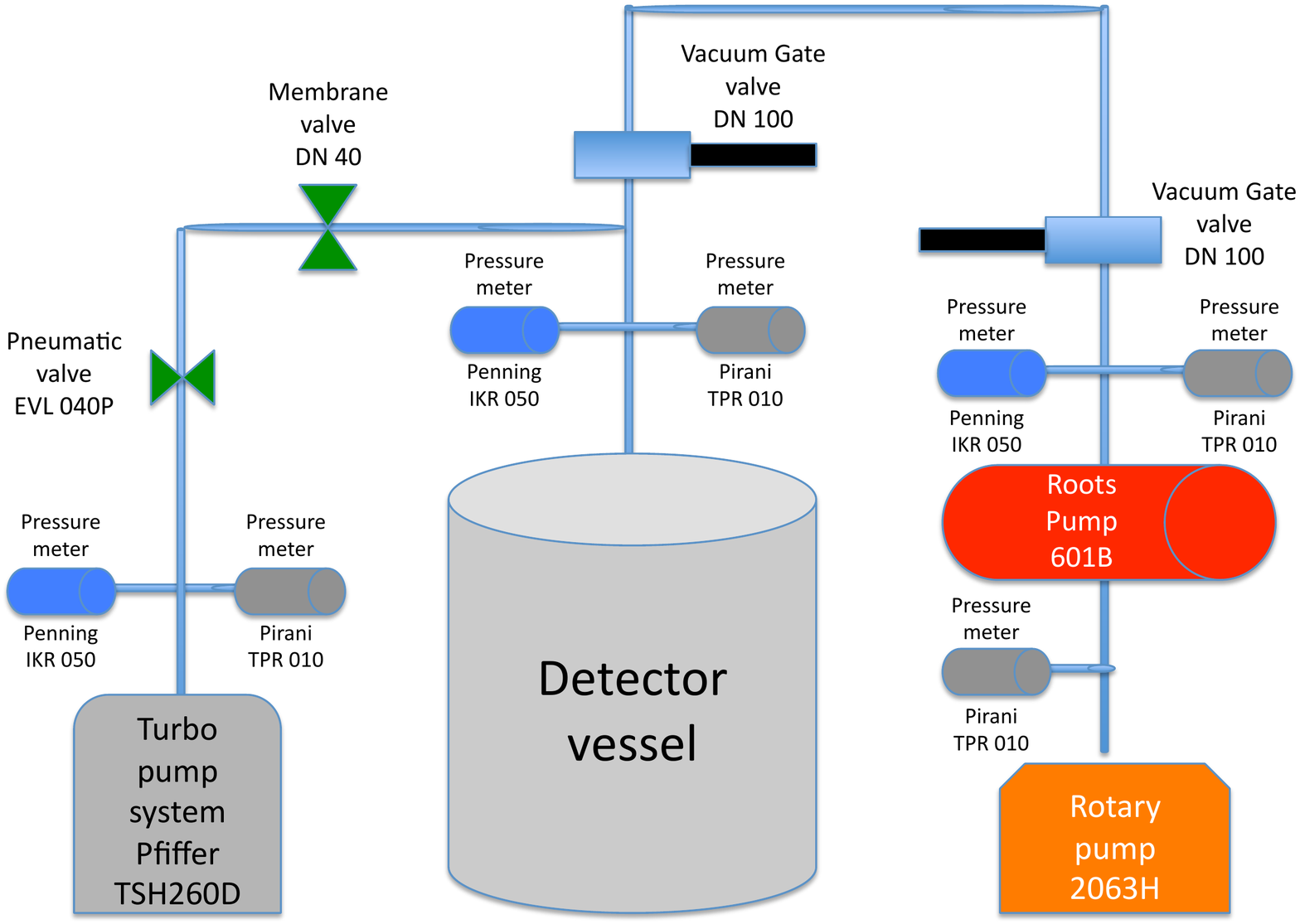}
\caption{The pumping system and its connection to the vessel.} 
\label{fig:pumps}
\end{figure}

Before filling with liquid Argon, the detector vessel is evacuated by means of a Roots pump\footnote{Alcatel RSV 601 B.} backed by a Rotary pump\footnote{Alcatel 2063 H.} and with a turbo molecular pumping system\footnote{Pfeiffer Vacuum TSH 260 D.} to reach a residual pressure of about 10$^{-4}$ mbar (figure~\ref{fig:pumps}). The cool down procedure of the apparatus starts by filling the external dewar with LAr. The reduction of temperature helps the pumping system reducing the residual pressure in a rather short time. Once 10$^{-5}$~mbar are reached, the pumping system is switched off.

The detector is then filled with ultra pure liquid Argon flown through the purification cartridge in liquid phase, analogous to the purification unit inserted in the re-circulation system shown in figure~\ref{fig:rec_pump}. The purity of the liquid Argon before passing through the OXYSORB cartridge has a nominal concentration of water and Oxygen of the order of the ppm. Filling of the TPC takes 5 hours and it is monitored by a capacitance\footnote{Cryo-Diffusion - Model INJC.} level metre made of a cylindrical capacitor about 1~m long. The increase of the liquid level in the vessel produces a directly proportional increase of the capacitance of the device.
%
\subsection{Data taking}
\label{sec:data_taking}
\indent

We did two campaigns of measurements with the detector continuously running for about one week, with the recirculation and purification system not yet in operation. During the first campaign the TPC was installed with the electrons drifting along the vertical direction. We performed both cosmic-ray and radioactive source (${}^{60}$Co) runs.  About 10000 cosmic-ray events and 5000 ${}^{60}$Co events were recorded. For the cosmic-ray runs the trigger was provided by the immersed PMT, while for the ${}^{60}$Co runs the trigger was made by  discriminating the signal of one wire of the collection plane. 

The second set of measurements was aimed at proving the feasibility of multi-photonic ionization of LAr. The TPC was installed with the electrons drifting along the horizontal direction (see Section~\ref{par:laser}). A UV laser run was successfully performed and more than 5000 ionization tracks were recorded for different values of the drift electric field. 
In addition, a second set of cosmic-ray and radioactive source (${}^{60}$Co) runs were performed.  About 5000 cosmic-ray and 50000 ${}^{60}$Co events were recorded in this case. 

Among the cosmic-ray data we found examples of passing through muons with multiple delta-rays, stopping and decaying muons, electromagnetic showers, hadronic interactions and electron positron pair production.

\begin{figure}
\center\includegraphics[width=14cm]{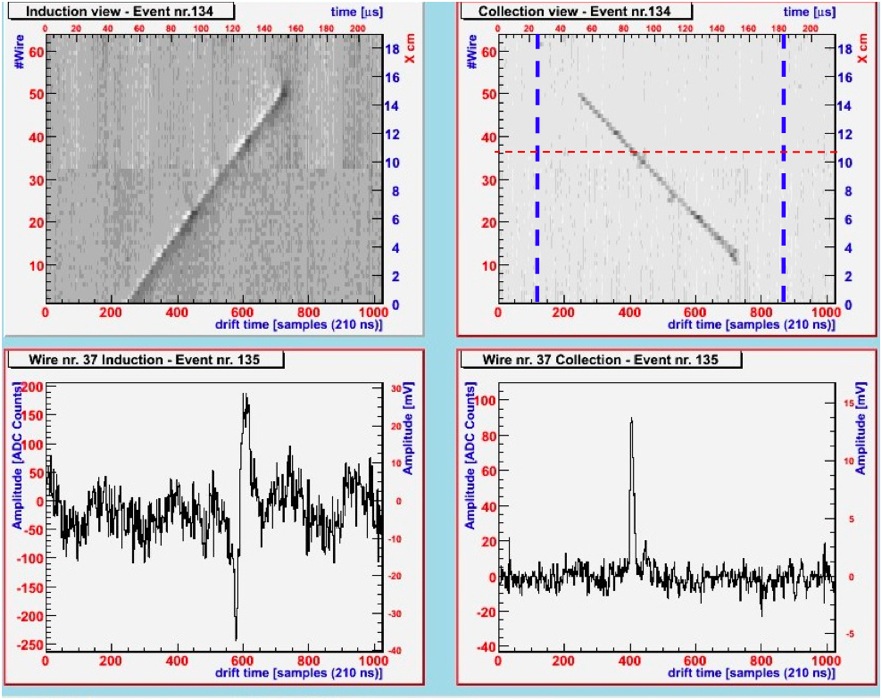}
\caption{Event display of a passing-through muon. The top-left box shows the induction view (X-Z projection). The top-right box indicates the collection view (Y-Z projection). Bottom boxes host the histograms of a single wire of the corresponding wire plane. For example, the bottom-right box shows the histogram of the wire pointed out by the red horizontal dotted line. The area between the dashed lines corresponds to the active volume of the TPC.}
\label{fig:first_event}
\end{figure}

\begin{figure}
\center\includegraphics[width=8cm]{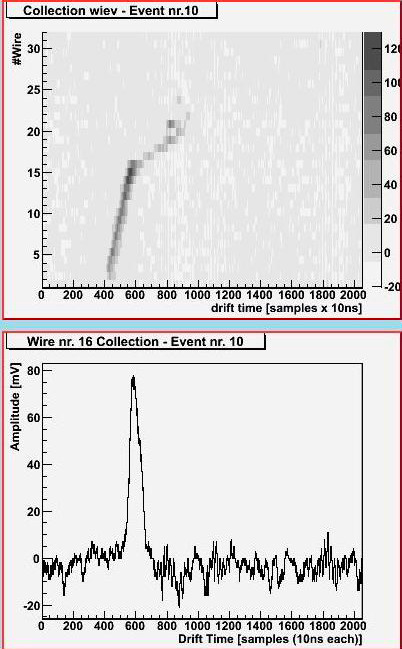}
\caption{Top:a muon decay event. Bottom: signal on the last wire hit by the muon.} 
\label{fig:muon_decay}
\end{figure}

\begin{figure}
\center\includegraphics[width=8cm]{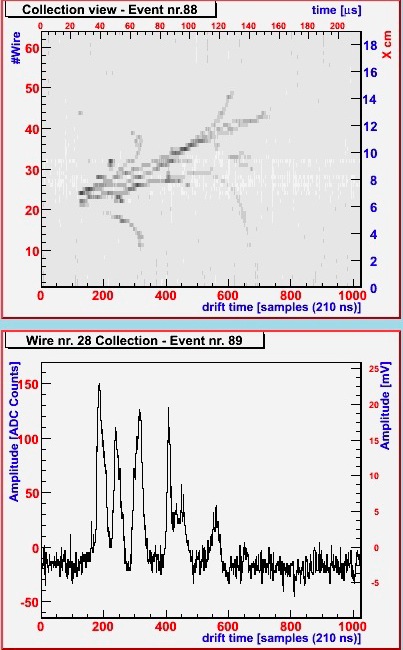}
\caption{"Shower" event induced by a cosmic-ray (Collection view). } 
\label{fig:e_magnetic}
\end{figure}

\begin{figure}
\center\includegraphics[width=8cm]{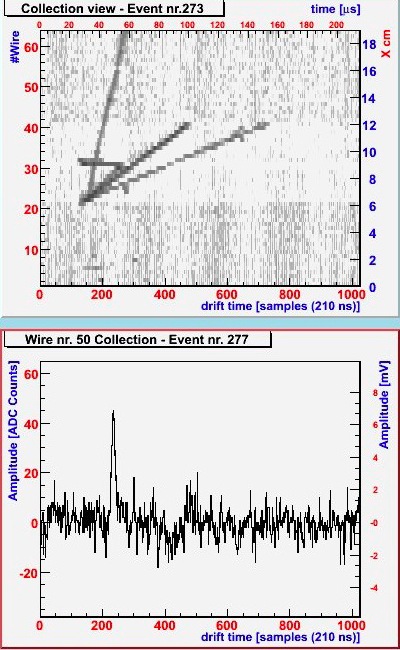}
\caption{Multi prong hadronic interaction (Collection view). } 
\label{fig:had_inter}
\end{figure}

\begin{figure}
\center\includegraphics[width=8cm]{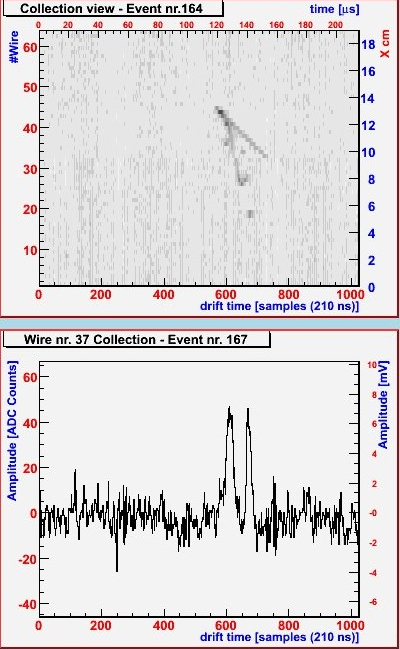}
\caption{Gamma conversion (pair production) event (Collection view). } 
\label{fig:pair}
\end{figure}

\begin{figure}
\center\includegraphics[width=8cm]{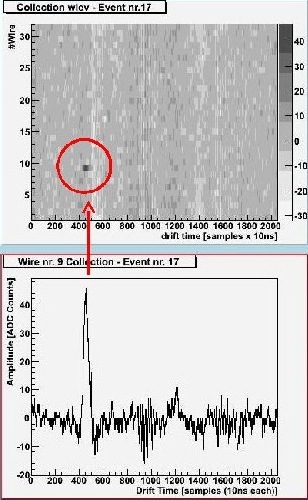}
\caption{Event display of a  typical ${}^{60}Co$ event (Collection view).} 
\label{fig:cobalt_event}
\end{figure}

\begin{figure}
\center\includegraphics[width=10cm]{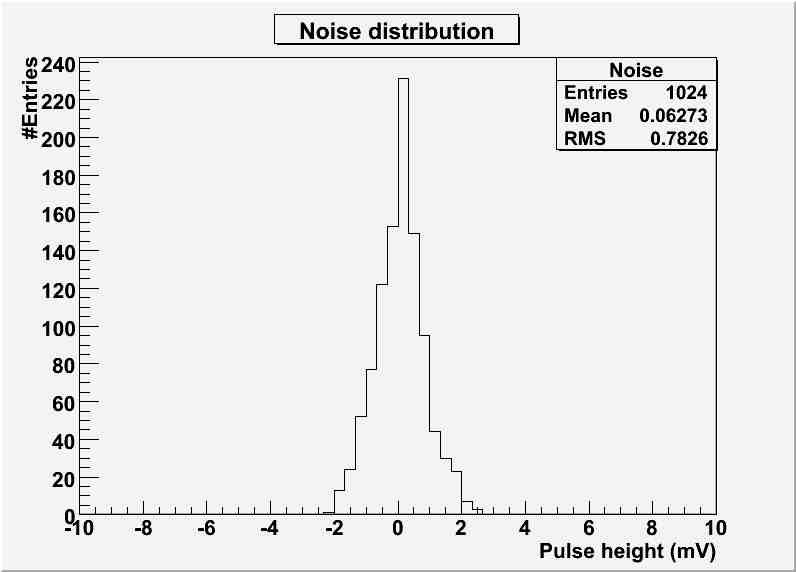}
\caption{Distribution of the peak value in mV for an empty collection wire, {\it i.e.} the electronic noise.} 
\label{fig:noise}
\end{figure}

Figure~\ref{fig:first_event} shows a passing-through muon event. Delta rays are visible along the muon track. Also shown is the waveform histogram of a single wire. In figure~\ref{fig:muon_decay} we depict the display of a muon decay event. As expected, the ionization yield increases along its path until the muon stops. The track of the decay electron is also visible. In figure~\ref{fig:e_magnetic} an electromagnetic shower is shown that extends over the entire fiducial volume. Several tracks are identified from the histograms of the single wire waveform. Figure~\ref{fig:had_inter} shows a hadronic interaction. The vertex is close to the wire planes. Four tracks with a high ionization density coming from the vertex are well visible.  Figure~\ref{fig:pair} shows a gamma conversion event with two short tracks emerging from the same vertex. The small amount of ionization indicates that the tracks are due to electrons. Figure~\ref{fig:cobalt_event} shows a ${}^{60}$Co event. The range of a Compton electron in LAr is of the order of few mm.
The average noise for the above measurements is about 0.8~mV (RMS), as shown in figure~\ref{fig:noise}, giving a signal-to-noise ratio of $\sim$8 for a minimum ionizing particle (mip). Given the wire pitch of 3~mm and an energy loss of 2.07~MeV/cm for a mip in LAr, the noise is roughly equivalent to $\sim$3000 electrons.

\begin{figure}
\center\includegraphics[width=10cm]{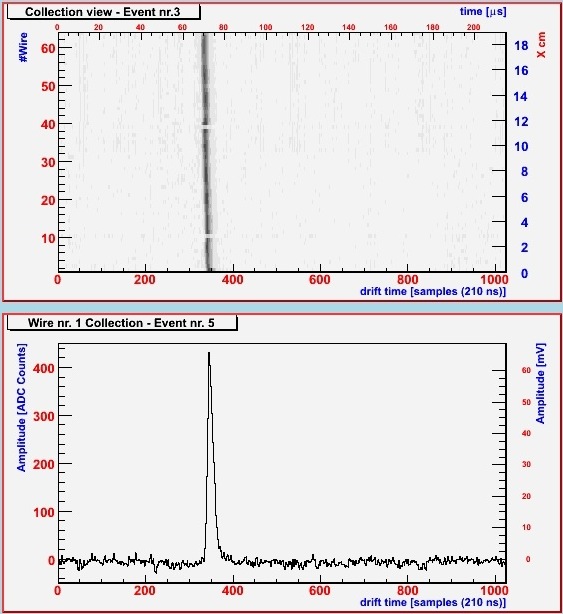}
\caption{Event display of a UV laser ionization track. The track generates the same signal over all the wires of the collection plane at the same time (drift coordinate). White spots along the laser track are caused by wires (\#10 and \#39) with lower gain due to a broken capacitance (Collection view).} 
\label{fig:laser_track}
\end{figure}

For the UV laser runs we recorded vertical laser tracks parallel to the wire planes (figure~\ref{fig:laser_track}) with a distance to the wire planes of 48~mm and an average energy of the laser of 20~mJ. 
The average noise for the UV laser measurements was the same as for cosmic-rays runs, giving a signal-to-noise ratio of $\sim$80 for 20~mJ beam energy. 


\subsection{Event reconstruction}
\label{par:reco}
\indent

A hit finding algorithm was developed specifically for the collection plane, which searched for hits independently for each wire. The algorithm first calculates the baseline for the waveform signals of each wire. A threshold is then set to $1.5 \sigma$ of the distribution of the entries recorded before the $t_0$ for each waveform. For each peak above the threshold a pulse width is defined as the group of consecutive samples above the 5\% of the peak value. 
If the pulse width is larger than~2~$\mu$s\footnote{This value is chosen based on the shaping times of the front-end preamplifiers.} a {\it hit candidate} is defined. For each candidate a fit is performed with a function 
\begin{equation}
f(t)=B+A\cdot \frac{e^\frac{-(t-t_{\mathrm{hit}})}{\tau_1} } {1+e^\frac{-(t-t_{\mathrm{hit}})}{\tau_2}}
\end{equation}
where B is the baseline, A is the amplitude, $t_\mathrm{hit}$ is the point for which the height of the function is equal to $A/2$, $\tau_1$ and $\tau_2$ are the falling and the rising characteristic times, respectively. If the fit parameters satisfy the following conditions: $A > 0$, $\tau_1>\tau_2$, $\tau_2\ge 0$ and $\tau_1\ge 0$, a hit is defined. The following informations are then stored: drift time ($t_{hit}$), pulse height, baseline value, pulse width and the integral of the fit over the pulse width (proportional to the charge released by the particle). No information from adjacent wires is used at this stage. 

A clusterization algorithm is performed for multiple hits in different wires. The algorithm associates hits belonging to a common charge deposition from adjacent wires in order to obtain a two-dimensional spatial reconstruction and the total energy associated to a track. Clusterization also provides criteria for particle ID and a more refined discrimination between noise and signal hits. 

All reconstructed hits have two degrees of freedom: the drift coordinate (common to both wire planes) and the wire number (specific to the plane). The redundancy in the drift coordinate permits the association of hits of the two different planes and a three-dimensional (3D) spatial reconstruction. Only hits from the collection plane provide information about the energy deposited by the particle.  

The charge associated to different hits is not directly comparable; infact, it depends on the drift coordinate (for a given purity) and on the angle between the track and the wire plane. In order to compare different hits the reconstructed energy is corrected by a {\it spatial} and a {\it purity} factor according to
\begin{equation}
E_{\mathrm{released/3mm}}=f_\mathrm{spatial}\cdot f_\mathrm{purity} \cdot E_{\mathrm{reco}}
\end{equation}
with $f_\mathrm{spatial}=p/\xi$, where $p$ is the wire pitch and $\xi$ is the lenght of the track subtended by the wire, and $f_\mathrm{purity}=e^\frac{(t-t_0)}{\tau}$. The parameter $\tau$ is the lifetime of the drifting electrons in LAr. In order to calculate $f_\mathrm{spatial}$ a 3D spatial reconstruction of the track is performed. 
%
\subsection{Events selection and results}
\label{par:results}
\indent

\begin{figure}
\center\includegraphics[width=14cm]{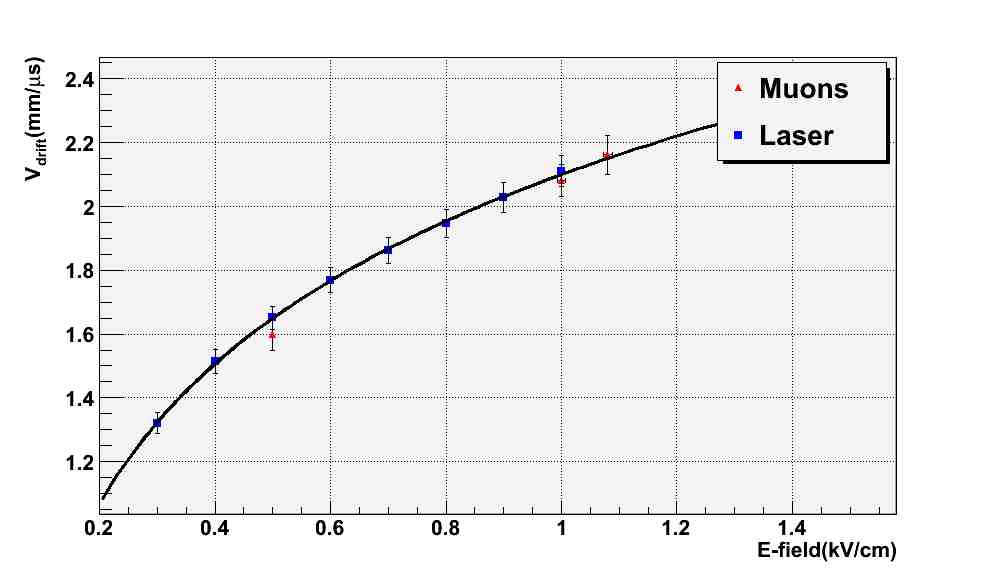}
\caption{Plot of the electron drift velocity in liquid Argon (87.3 K) as a  function of the applied electric field. Blue squares indicate the measurements done with UV laser, while red triangles represent measurements done with long minimum ionizing tracks crossing the volume from anode to cathode. The fit of the laser data was performed with an empirical formula taken from \cite{[Kalinin], [Walkowiak]}
}
\label{fig:e_velocity}
\end{figure}

The velocity of the drifting electrons was measured for different values of the electric field with cosmic muons and with UV laser beams. For the measurements with muons we selected events with almost vertical track passing through the wire planes and the cathode. We determined the velocity from the time interval between the closest hit to the wire plane (the $t_0$ of the PMT) and the closest hit to the cathode and the a priori information of the maximum drift length (263$\pm$1~mm). For the measurements with the UV laser, the beam was  vertically steered in the active volume of the TPC. The centre of the beam had a distance of 48$\pm$1~mm to the wire planes. Consequently, each laser pulse had almost the same drift time for all collection wires, as can be seen in figure~\ref{fig:laser_track}. The time interval between the $t_0$ (given by the laser itself) and the hits provided by the laser track were measured. The precision of this measurement is related to the ADC sampling, which for these data sample was set equal to 210~ns. The results are summarized in table~\ref{tab:speed} and in figure~\ref{fig:e_velocity}. The data agree well with the theoretical predictions and previous results \cite{[Walkowiak], [purity]}.
\begin{table}[ht]
\caption{Velocity of drifting electrons for different $E_{field}$ (T=87.3~K).}
\centering
\begin{tabular}{ c | c c }
\hline\hline
& $E_{field}$~(kV/cm)&  v~(mm/$\mu$s)   \\
\hline
\multirow{3}{*}{\rotatebox{90}{\small{Muons}}} 
&$ 1.08 $  & $2.16\pm 0.06$   \\
&$ 1.00 $  & $2.08\pm 0.05$   \\
&$ 0.50 $  & $1.60\pm 0.05$   \\
\hline
\multirow{8}{*}{\rotatebox{90}{Laser}}
&$ 1.00 $  & $2.11\pm 0.05$   \\
&$ 0.90 $  & $2.03\pm 0.05$   \\
&$ 0.80 $  & $1.95\pm 0.04$   \\
&$ 0.70 $  & $1.86\pm 0.04$   \\
&$ 0.60 $  & $1.77\pm 0.04$   \\
&$ 0.50 $  & $1.65\pm 0.04$   \\
&$ 0.40 $  & $1.51\pm 0.04$   \\
&$ 0.30 $  & $1.32\pm 0.03$   \\
\hline	
\end{tabular}	
\label{tab:speed}
\end{table}

\begin{figure}
\center\includegraphics[width=14cm]{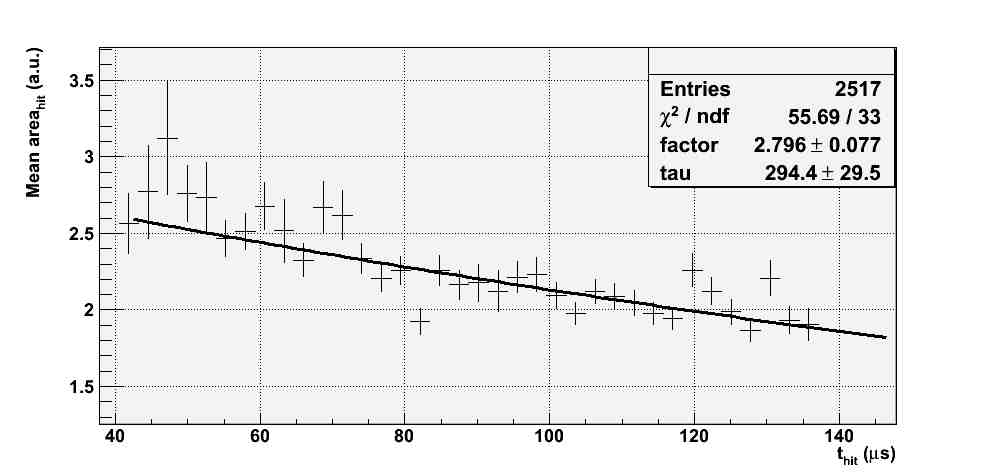}
\caption{Profile histogram of the hit area (corrected by the spatial factor) as a function of the drift time. The exponential fit allows to determine the drifting electrons lifetime in liquid Argon for this set of data.}
\label{fig:purity}
\end{figure}

Fig.~\ref{fig:purity} shows the results of a measurement of the drifting electrons lifetime. A set of 180 passing-through muons with a large vertical angle was used. A manual 3D reconstruction of the data set was performed in order to obtain the spatial correction factor. The plot in the figure is the profile histogram of the hit area (proportional to the charge released) as a function of the drift distance of the waveform peak. The range on the abscissa is given by the total drift length of the TPC. The $t_0$, provided by the PMT and fixed by the DAQ hardware settings at 42~$\mu$s, represents the position of the wire planes. In the case of 100\% purity, one would expect a flat curve. Therefore, any deviation provides a measurement of the actual purity.

The curve on the figure was fit with an exponential function $f(t)=A\cdot e^{-(t_{hit}-t_0)/\tau}$ where $t_0$ is the position (along the abscissa) of the wire planes, $t$ is the drift coordinate and $\tau$ is the drifting electrons lifetime in LAr. The fit yields a lifetime $\tau=0.29\pm$0.03~ms\footnote{The error of the purity measurement comes from the errors of the parameters of the fit function.}, corresponding to an attenuation length $\lambda$ = 600~mm at 1~kV/cm\footnote{$\lambda=v_d \tau$, where $v_d$ is the drift velocity.}.
%
\section {Conclusions}
We reported on the design, realization and operation of a prototype liquid Argon Time Projection Chamber detector and on the development of a novel online monitoring and calibration method exploiting ionizing UV laser beams. 

We described the experimental methods and the design of a setup used for a first set of measurements performed at the University of Bern. Data were taken with cosmic-rays, a radioactive ${}^{60}$Co source and beams from the $4^{\mathrm{th}}$ harmonic of a Nd-YAG laser. 

In the first campaign of measurements, the detector was continuously running for about one week. About 10000 cosmic-ray events and 5000 ${}^{60}$Co events were recorded and analyzed.
In the second set of measurements, UV laser runs were successfully performed and more than 5000 ionization tracks were recorded together with additional cosmic-ray and ${}^{60}$Co events. 
The detailed analysis of these data is in progress and will be the subject of a forthcoming publication. 
 
The results obtained and described in this paper make us confident on the practical use of UV laser beams for the calibration of LAr TPC detectors and for the monitoring of the LAr purity, as requested for the envisioned large scale applications of the technique.
%
%
\section* {Acknowledgements}
The work presented in this paper was conducted thanks to grants from the Swiss National Science Foundation (SNF) and from the University of Bern. We warmly acknowledge both institutions.

\end{document}